\documentclass[aps,prb,twocolumn,groupedaddress]{revtex4}
\usepackage{graphicx}
\usepackage{amssymb}
\usepackage{amsmath}
\usepackage{bm}

\newcommand{\uu}{\uparrow}
\newcommand{\dd}{\downarrow}

\newcommand{\ket}[1]{|{#1}\rangle}
\newcommand{\bra}[1]{\langle{#1}|}
\newcommand{\abs}[1]{\left|{#1}\right|}

\newcommand{\Heff}{H_\textrm{eff}}

\begin{document}

\title{Symmetries, Topological Phases and Bound States in the One-Dimensional
  Quantum Walk}
\author{J. K. Asb\'oth}
\affiliation{
Institute for Solid State Physics and Optics, Wigner Research Centre, Hungarian Academy of Sciences, H-1525 Budapest P.O. Box 49, Hungary}
\date{Autumn 2012}
\begin{abstract}
  Discrete-time quantum walks have been shown to simulate all known
  topological phases in one and two dimensions. Being periodically
  driven quantum systems, their topological description, however, is
  more complex than that of closed Hamiltonian systems. We map out the
  topological phases of the particle-hole symmetric one-dimensional
  discrete-time quantum walk. We find that there is no chiral symmetry
  in this system: its topology arises from the particle--hole symmetry
  alone. We calculate the $\mathbb{Z}_2\times\mathbb{Z}_2$ topological
  invariant in a simple way that is consistent with a general
  definition for 1-dimensional periodically driven quantum
  systems. These results allow for a transparent interpretation of the
  edge states on a finite lattice via the the bulk--boundary
  correspondance. We find that the bulk Floquet operator does not
  contain all the information needed for the topological invariant. As
  an illustration to this statement, we show that in the split-step
  quantum walk, the edges between two bulks with the same Floquet
  operator can host topologically protected edge states.
\end{abstract}
\maketitle

\section{Introduction}

The quantum mechanical generalization of the random walk has, since
its first definition \cite{aharonov93} received quite some interest.
Its hallmark property is that as opposed to the classical walk, the
standard deviation of the position of the walker increases linearly with
time. This $\sqrt{t}$ speedup over the classical diffusive scaling
lies behind the advantage of the Grover search.
This is all the more interesting, as a variant of the quantum walk can realize a general purpose 
quantum computer \cite{childs_universal, childs_arxiv}.  Quantum walks have also 
attracted attention as a convenient platform to study the effects of 
decoherence \cite{kendon_review}. 
The surge of interest in quantum walks has resulted in their
experimental realization in varied physical systems, such as trapped ions
\cite{roos_ions,schmitz_ion}, cold atoms in optical
lattices\cite{meschede_science}, and on photons on an optical table\cite{gabris_prl,white_photon_prl}.


 

A discrete time, coined quantum walk can be viewed as a stroboscopic
simulation of time evolution by an effective Hamiltonian.  The
topological features of lattice
Hamiltonians has in the last decade been the focus of intense interest
in solid state physics.  The so-called bulk--boundary correspondence,
showing how differences between bulk topologies give rise to
low-energy states residing at the ``edges'', the boundaries between
these bulks, is at the heart of the general theory of topological
insulators \cite{rmp_zhang, rmp_kane}.  Recently, Kitagawa et al. have
shown how, by varying the parameters of the discrete-time quantum
walk, one can realize all known kinds of topological phases in 1 and 2
dimensions \cite{kitagawa_exploring, kitagawa_introduction}. The
striking physical consequence is that in an inhomogeneous system, a
walker started at a boundary between domains with different topology
can be localized (1D case), or propagate unidirectionally (2D
case)\cite{kitagawa_exploring}. This ``trapping effect'' for a 1D
quantum walk has already been seen in an experiment performed with
photons \cite{kitagawa_observation}. The appearance of these
``trapping states'' in a disordered quantum walk can lead to
sub-diffusive spreading of the
wavefunction\cite{obuse_delocalization,gabris_prl}, a phenomenon familiar from
disordered superconducting wires \cite{brouwer_delocalization}.

In the lab, a discrete-time quantum walk is realized by
periodically modulating the parameters of the experimental
setup. Compared to a closed, time-independent system, a quantum walk
thus can have a broader range of ways in
which topology can enter its description.  
One example is the winding of quasienergy\cite{kitagawa_periodic},
which can lead to novel kinds of edge states. However, even for the
quantum walks where the winding of the quasienergy is 0, it is a
relatively unexplored question to what extent their topological
properties go beyond that of the underlying effective Hamiltonian.

The bulk-boundary correspondence predicts ``edges states'' where the edges
are defined by assigning a position dependence to the parameters of a
system. In many practical situtations, however, an edge represents the
physical boundary of the system.  For Hamiltonian systems, the
simplest approach, called ``open boundary conditions'', is to set the
hopping rates to zero at the edge.  For quantum walks, boundaries are
realized by using reflective coins, or -- the analogue of open
boundary conditions -- cutting the links. Topologically protected
states at such boundaries have been predicted
\cite{oka_breakdown,obuse_delocalization}, and analyzed using an
adiabatic argument\cite{kitagawa_introduction}. However, their
relation to the bulk-boundary correspondence is so far not understood.

In this paper, we revisit the question of the topological phases of
the 1-dimensional discrete-time quantum walk.  In Section
\ref{sect:definition} we define the quantum walk that we are going to
study, along with the introduction of the associated effective
Hamiltonian. In Section \ref{sect:symmetries} we analyze the
symmetries of the system. Our choice of coin operator, which is widely
used \cite{kitagawa_exploring,
  kitagawa_introduction,obuse_delocalization} ensures Particle-Hole
Symmetry of the effective Hamiltonian. However, contrary to previous
works \cite{kitagawa_exploring, kitagawa_introduction,
  obuse_delocalization}, we find that there is no chiral symmetry for
this walk. We also discuss the sublattice symmetry of the time
evolution operator: this turns out to cause energy eigenstate to come
in pairs, but otherwise have no significant consequences for the
topology. In Section \ref{sect:topological_phases} we explore the
topological phases of the quantum walk. At variance with Kitagawa et
al.\cite{kitagawa_exploring}, we find two different topological phases
for the simple 1D quantum walk. A spatial boundary between domains
with different topology hosts a pair of topologically protected bound
states. We show that a naive way to determine the relative values of
these invariants is in line with the definition of the topological
invariant for periodically driven quantum systems due to Jiang et
al. \cite{akhmerov_majorana}.

In Section \ref{sect:finite_line} we consider the quantum walk on a
finite line. Termination of the lattice by a completely reflective
coin operator and ``open boundary conditions'' by cutting the links
have already been considered, but we rederive the results using the
bulk-boundary correspondence for completeness. Cutting the links at the boundary leads us to a generalization of the
discrete time quantum walk which is equivalent to the split-step
walk \cite{kitagawa_exploring}. We find that the split-step walk has a
$\mathbb{Z}_2\times \mathbb{Z}_2$ topological invariant, which is
unique to periodically driven quantum systems. We map out the
parameter space of the split-step walk. This allows us to predict that
a generic 1D particle-hole-symmetric discrete time quantum walk has a
single topologically protected edge state at each ``open boundary'',
with energy $E=0$ or $E=\pi$, depending on the topology of the bulk
and on how the link at the boundary is cut. This is in contrast to
boundaries defined by reflective coins, where either a pair of bound
states with energies $E=0$ and $E=\pi$ are present, or no bound states
at all. Finally, we provide a striking example of the way in which
periodically driven systems have topological features not present in
their effective lattice Hamiltonians: A boundary between two quantum
walks with the \emph{same} bulk timestep operator supporting a pair of
edge states with energies $E=0$ and $E=\pm\pi$.

\section{Discrete time quantum walk}  
\label{sect:definition} 
The quantum walk we consider in this paper is a standard extension of
the common discrete-time quantum walk.
We consider a particle with a discrete position degree of freedom, $x =
0,\ldots,N$, and two internal (coin) states, labeled $\uparrow$ and
$\downarrow$. Thus, the quantum state of the particle can be represented by a
complex $2N$-component vector:
\begin{align}
\ket{\Psi} = \sum_{x=1}^{N} \left( \Psi_{x,\uu}  \ket{x}\otimes\ket{\uparrow} +
  \Psi_{x,\dd} \ket{x}\otimes\ket{\downarrow} \right).
\end{align}

\begin{figure}
\includegraphics[width=8cm, trim=0.5cm 0.5cm 0.4cm 0.cm, clip=true]{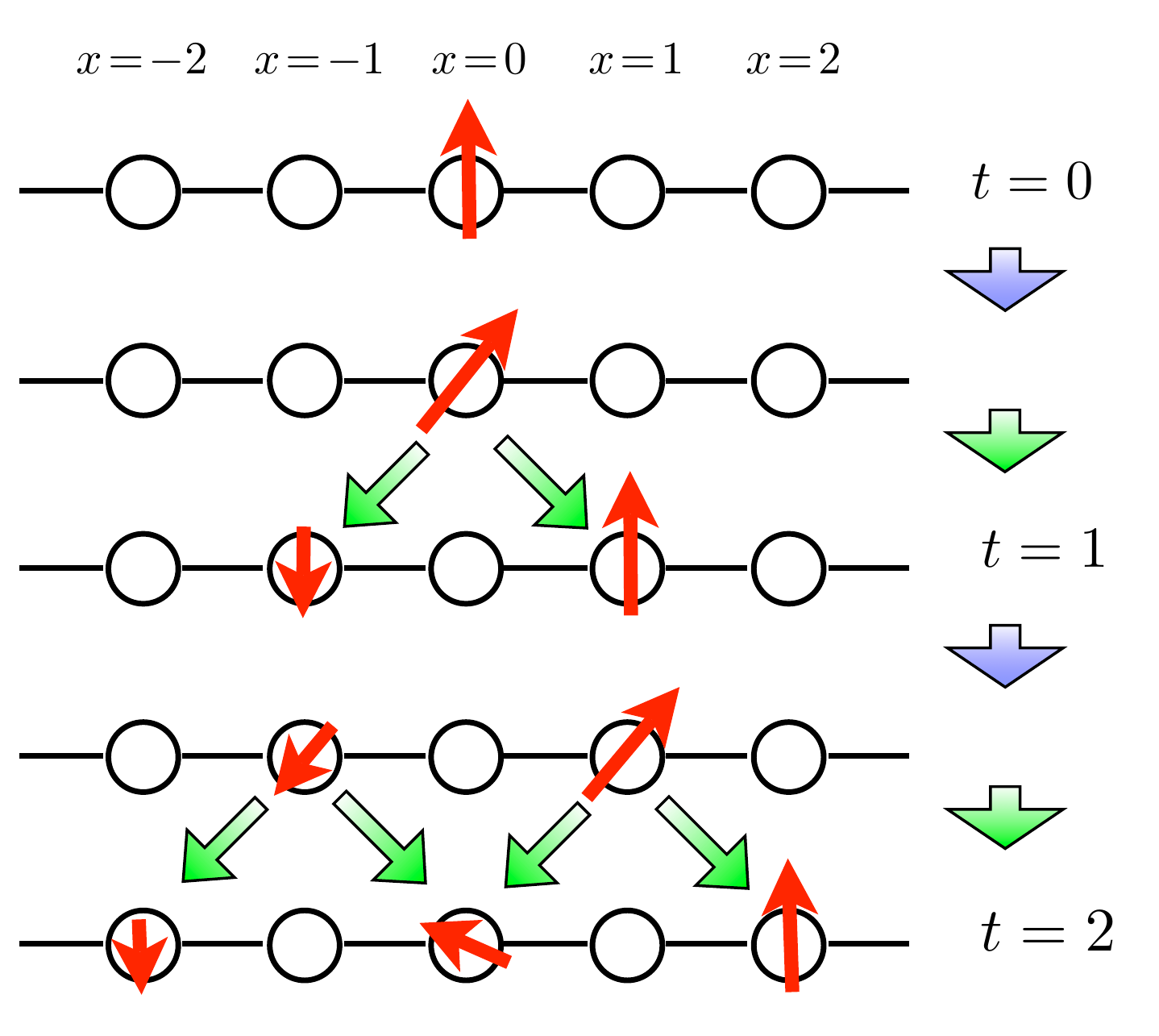}
\caption{The discrete-time quantum walk. A spin-$1/2$ particle
  starting from a site of a discrete lattice undergoes alternating
  spin rotations $R$ and spin-$z$ dependent unitary shifts $S$. The
  first few timesteps are shown representing the effect of
  interference.   
} 
\label{fig:quantum_walk_illustrate}
\end{figure}

The dynamics of the quantum walk is given by a unitary timestep
(Floquet) operator, consisting of a rotation of the spin followed by a
spin-dependent shift of the particle,
 \begin{align} 
\ket{\Psi(t+1)} &= U \ket{\Psi(t)} = S R \ket{\Psi(t)}.
\label{eq:U_def}
\end{align} 
This is illustrated in Fig.~\ref{fig:quantum_walk_illustrate}.
Conveniently, we choose the unit of time to be the period of the time
evolution, the unit of position the period of the lattice, and set $\hbar=1$.

The operator $S$ translates the particle by one lattice site to the
left (right), if its spin is pointing down(up),
\begin{align} 
S &= \sum_{x=1}^{N}\big( \ket{x-1}\bra{x} \otimes \ket\dd\bra\dd + 
\ket{x+1}\bra{x} \otimes \ket\uu\bra\uu \big).
\label{eq:S_def}
\end{align} 
Periodic boundary conditions are taken, i.e., $N+1=1$.  For a
translation independent bulk, we use the Fourier transform, $\ket{k} =
\frac{1}{\sqrt{N}} \sum_x e^{-ikx} \ket{x}$, and can write the
particle shift operator as
\begin{align}
S &= \sum_k \left\{ e^{-ik} \ket\uu \bra\uu + e^{ik} \ket\dd \bra\dd
\right\}\otimes\ket{k}\bra{k} = e^{-ik\sigma_z}.
\end{align}
Here and in the following the operators $\sigma_{x,y,z}$ denote the
Pauli matrices acting on the internal ``pseudospin'' degree of
freedom, with basis states $\ket\uparrow, \ket\downarrow$. 

The coin operator $R$ is a unitary rotation in the internal space of 
the particle (corresponding to the
``coin flip'' in the classical walk.) It is diagonal in $x$,
\begin{align}   
R &= \sum_{x} \ket{x}\bra{x} \otimes R(x).
\end{align}   
We require $[R(x),R(x')]=0$ and $\sigma_zR(x)\sigma_z=R(x)^{-1}$ for every
$x,x'$, in order to 
ensure particle-hole symmetry (see details later). In that case, 
without any loss of generality, we can take $R(x)$ to be a unitary rotation 
of the spin around the $y$ axis by a position-dependent angle $\theta$, 
\begin{align}   
R(x) = R(\theta(x)) = \exp(-i \theta(x) \sigma_y).
\label{eq:R_def}
\end{align}

\subsection{Effective Hamiltonian}

To realize the quantum walk, we need an experimental setup with
time-dependent external fields. Denoting the explicitly time-dependent
Hamiltonian by $H(t)$, we have 
\begin{align}
U=\mathbb{T}e^{-i\int^1
  H(t) dt},
\end{align}
where $\mathbb{T}$ is the time-ordering operator.  Taking the
logarithm of $U$, we can associate a time-independent effective
Hamiltonian $\Heff$ to this unitary operator (cf. Floquet theory),
defined as
\begin{align}  
U &= e^{-i \Heff}.
\label{eq:Heff_def}
\end{align}  
In the translation invariant bulk, the time evolution operator is
diagonal in momentum space, $U = \sum_k U(k) \otimes \ket{k}\bra{k}$, with 
\begin{align}  
U(k) &= e^{-ik\sigma_z} e^{-i\theta \sigma_y} = e^{-i \Heff(k)}.
\label{eq:Uk_def}
\end{align}  

In the bulk, the quantum walk realized by $H(t)$
stroboscopically simulates the time evolution via $\Heff$.  The
eigenvalues of the effective Hamiltonian $\Heff$ are the
quasienergies, which can be restricted to an energy Brillouin zone
$-\pi,\ldots, \pi$, in the same way as the quasimomenta are restricted
to the first Brillouin zone.  Since $U$ is a product of SU(2)
operators, its determinant is 1, thus $\Heff$ has to be traceless, and
the spectrum has to be symmetric around $E=0$. Note that this is a
property of the spectrum and not of $\Heff(k)$, and in itself implies
neither particle-hole symmetry (ensured by our choice of $R$) nor
chiral symmetry (absent in this system: see Section \ref{sect:nochiral}) of the effective Hamiltonian
$\Heff$. However, it does mean that there can be no winding in
quasienergy \cite{kitagawa_periodic}.



\section{Symmetries and gaps}
\label{sect:symmetries}

To understand what topological phases and topologically protected edge
states the quantum walk might have, we need to examine the
symmetries and the related protected gaps of the effective Hamiltonian.  
 
\subsection{Particle-Hole Symmetry}

The unitary timestep operator (cf Eqs.~\eqref{eq:U_def},
\eqref{eq:S_def}, \eqref{eq:R_def}) in position and $\sigma_z$-basis has only real elements: 
\begin{align}
      U^\ast &= U,
\end{align}
where here and in the following $*$ denotes complex conjugation in the
$x$ and $\sigma_z$-basis. 
By the definition of the effective Hamiltonian, this implies
\begin{align}
      \Heff^\ast &= -\Heff; \quad\Longrightarrow \Heff^\ast(-k) = -\Heff(k)
\end{align}
For stationary states $\ket{\Psi}$ of the walk this translates to
\begin{align*}
  \Heff \ket{\Psi}^\ast &= -E \ket{\Psi}^\ast 
\end{align*}

Thus we have Particle-Hole Symmetry (PHS), with $P^2 = 1$. It is
represented by complex conjugation: $E\leftrightarrow-E;
\ket{\Psi}\leftrightarrow \ket{\Psi}^\ast$ \cite{kitagawa_exploring}.


Eigenstates of the quantum walk with energy $0$ or $\pi$ can be their
own particle-hole symmetric partners -- this happens if their
wavefunctions are real. If there is a bulk gap around these states (if
these are midgap states), their energies can be protected
against particle-hole symmetric perturbations. 

\subsection{Sublattice Symmetry}

The lattice on which the walk takes place is bipartite: we can assign
each lattice site $j$ to one of the sublattices $\alpha$ and $\beta$,
with every link connecting sites from different sublattices. Moreover,
the lattice of the unitary timestep operator itself is bipartite:
\begin{align}
  U &= \underset{\langle jl\rangle}{\sum}  
U_{jl} \ket{j}\bra{l} + U_{lj} \ket{l}\bra{j} \,\,: \,j\in \alpha;\quad l\in \beta, 
 \end{align}
where the $U_{jl} = \bra{j} U \ket{l}$ are operators in spin space.
This leads to a symmetry of the effective
Hamiltonian\cite{obuse_delocalization}, that is sometimes called
``chiral symmetry''\cite{shikano_pre}.  Since this symmetry arises
from the bipartition of the timestep operator, we are going to call it
``sublattice symmetry''.

Defining the sublattice
operator $\tau_z$, we can express sublattice symmetry in a concise way:
\begin{align}
\label{eq:tau_z_def}
 \tau_z &\equiv \underset{j\in \alpha}{\sum} \ket{j}\bra{j}
- \underset{l\in \beta}{\sum} \ket{l} \bra{l};\\
\tau_z U \tau_z &= -U. 
\end{align}
Substituting the definiton of $\Heff$ from Eq.~\eqref{eq:Heff_def}, we
obtain
\begin{align}
  \tau_z \Heff \tau_z &= \Heff+\pi  .
\end{align}
For energy eigenstates $\ket{\Psi}$, this means
\begin{align}
  \Heff \tau_z\ket{\Psi} &= (E+\pi) \tau_z\ket{\Psi} 
\end{align}
 
Note that $\tau_z$ is a local operator: we can extend the unit cell in
such a way that the matrix of $\tau_z$ is translation invariant, and
does not link different unit cells. Moreover, $\tau_z$ is independent
of all of the angles $\theta(x)$, and so defines a unitary symmetry
for the whole set of Hamiltonians $\{\Heff(\theta(x))\}$.

Sublattice symmetry (SLS) does not change the number of independent,
symmetry protected gaps. On the one hand, SLS implies that the bulk
has a gap around $E=\pi$ if and only if it has a gap around $E=0$:
This decreases the number of independent, symmetry protected gaps from
2 to 1. On the other hand, however, there is a new kind of protected
gap. For a state with with energy $\pi/2$, its SLS partner can
coincide with its PHS partner.  This happens, e.g., if the
wavefunction is real on even and imaginary on odd sites.  Assuming
there is a bulk gap around energy $\pi/2$ (and therefore around
$E=-\pi/2$ as well), the energies of this pair of states are protected
by SLS and PHS.

\subsection{No chiral symmetry}  
\label{sect:nochiral}
Importantly, it is the lattice of the timestep operator $U$, and not
of the effective Hamiltonian $\Heff$, that is bipartite. If the
Hamiltonian was bipartite, that would give us chiral symmetry, with a
unitary operator $W = \tau_z$, as defined in Eq.~\eqref{eq:tau_z_def}, and
\begin{align}
W \Heff W^\dagger = - \Heff. 
\end{align} 
Here, we find no local unitary operator $W$ representing such a symmetry. 

Kitagawa et al. \cite{kitagawa_exploring} identify a ``chiral
symmetry'' for the system, with $W = \cos \theta \sigma_x + \sin
\theta \sigma_z$. However, since this operator depends explicitly on
$\theta$, we do not think it should be considered a ``symmetry''.
Whenever symmetry properties of a system are investigated, it is not
only one specific Hamiltonian, but an ensemble of Hamiltonians that
should be considered. The operator representing the symmetry has to be
the same for all elements of the ensemble. The ensemble we consider
here, are the quantum walks with varying rotation angles
$\theta$. This follows from the fact that $\theta$ is the only tunable
parameter of the walk that we can use, e.g., to create an
inhomogeneous system with different domains. Since the ``chiral
symmetry operator'' $W$ depends explicitly on $\theta$, it does not
represent a symmetry of the system.  (In an inhomogeneous system,
$\theta$ is a spatially varying parameter, and so $W$ is not even
properly defined.)

Since $U(k)$ has determinant 1, $\Heff(\theta,k)$ is traceless, and
therefore its spectrum is symmetric for any $\theta$. This could hint
at chiral symmetry: a unitary operator $W$ that transforms the
positive energy eigenstate $\ket{+,\theta,k}$ of $\Heff(\theta,k)$
into its negative energy eigenstate, $\ket{-,\theta,k}$, and vice
versa.  However, for different values of $\theta$, as $k$ is swept
through $[-\pi,\pi]$, the eigenstate $\ket{+,\theta,k}$ takes on every
value on different great circles on the Bloch sphere
\cite{kitagawa_exploring}.  Unitary transformations are rotations on
the Bloch sphere, and there is no rotation that takes every point to
its antipodal pair on two different great circles.  Therefore there is
no chiral symmetry for the effective Hamiltonian of the discrete time
quantum walk. Since we have particle-hole symmetry, the absence of
chiral symmetry also precludes the existence of time reversal
symmetry of the effective Hamiltonian.

\section{Topological Phases of the Quantum Walk}
\label{sect:topological_phases}
\begin{figure}
\includegraphics[width=8cm, trim=0.5cm 0.5cm 0.4cm 0.cm, clip=true]{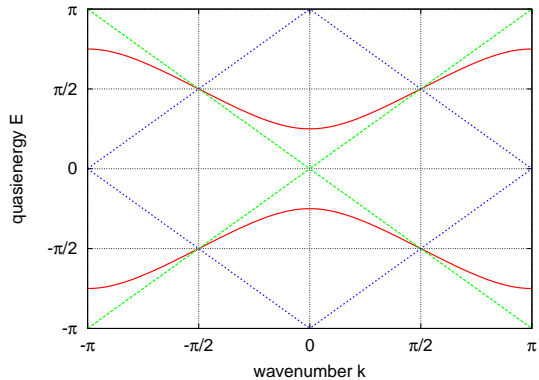}
\caption{Dispersion relations of the 1D quantum walk.
Continuous line shows a typical gapped phase, with $\theta =
\pm\pi/4$. The two gapless dispersion relations are: $\theta = 0$
(slashed) and $\theta=\pm \pi$ (dotted). In both cases the gaps at
$E=0$ and $E=\pm\pi$ are closed.  
} 
\label{fig:dispersions}
\end{figure}

\begin{figure}
    \includegraphics[width=8cm]{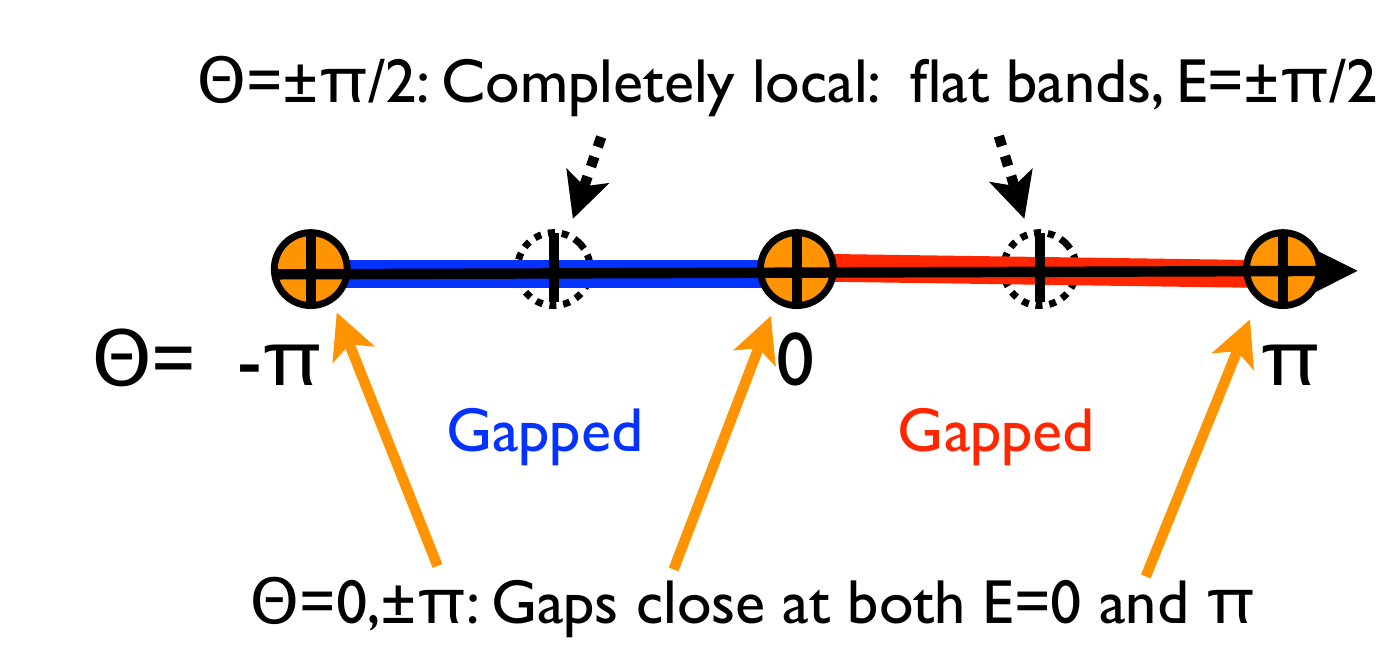}
\caption{Parameter space of the 1-dimensional simple quantum walk. The
  only parameter is the coin rotation angle $\theta$. The parameter
  space consists of two gapped domains, with gaps around both $E=0$
  and $E=\pi$. These are separated by the gapless
  points, $\theta=0$, and $\theta=\pi$. 
} 
\label{fig:parameter_space}
\end{figure}

To understand the topological phases of the quantum walk, we treat
the translation independent (bulk) case, i.e.,  
$\theta(x)=\theta$ independent 
of $x$. 
The dispersion relation of the effective Hamiltonian follows from
eq.~\eqref{eq:Uk_def} in a straightforward
way\cite{kitagawa_exploring},
\begin{align}
\label{eq:disprel_1}
\cos E(k) &= \cos (k) \cos (\theta).
\end{align}
The resulting dispersion relations for generic values of $\theta$, and
for the special values $\theta=0$, and $\theta=\pi$ are plotted in 
Fig.~\ref{fig:dispersions}. Note that for generic rotation angle
$\theta$, the dispersion relation has gaps around $E=0$ and around
$E=\pi$. 

At the time reversal invariant momenta $k=0$ and $k=\pi$, the Floquet
operator $U(k)$, as in Eq. \eqref{eq:Uk_def}, has a particularly
simple form: 
\begin{align}
U(k=0)&=e^{-i\theta \sigma_y},\\
U(k=\pi)&=e^{-i(\pi+\theta \sigma_y)}.
\end{align}
This shows directly that the dispersion relation has gaps at $k=0, E=0$ and at $k=\pi,
E=\pi$ of magnitude $\theta$. Thus, the parameter space
$\theta=\pi,\ldots,\pi$ falls apart to two disconnencted intervals
where the system is gapped: $-\pi<\theta<0$ and $0<\theta<\pi$.
This is illustrated in Fig.~\ref{fig:parameter_space}.
This allows for the possibility that these regimes correspond to two
distinct topological phases.  

\subsection{Edge states in the simple quantum walk}  
\begin{figure}
\includegraphics[width=8cm, trim=3.1cm 4.5cm 1.4cm 4.0cm, clip=true]{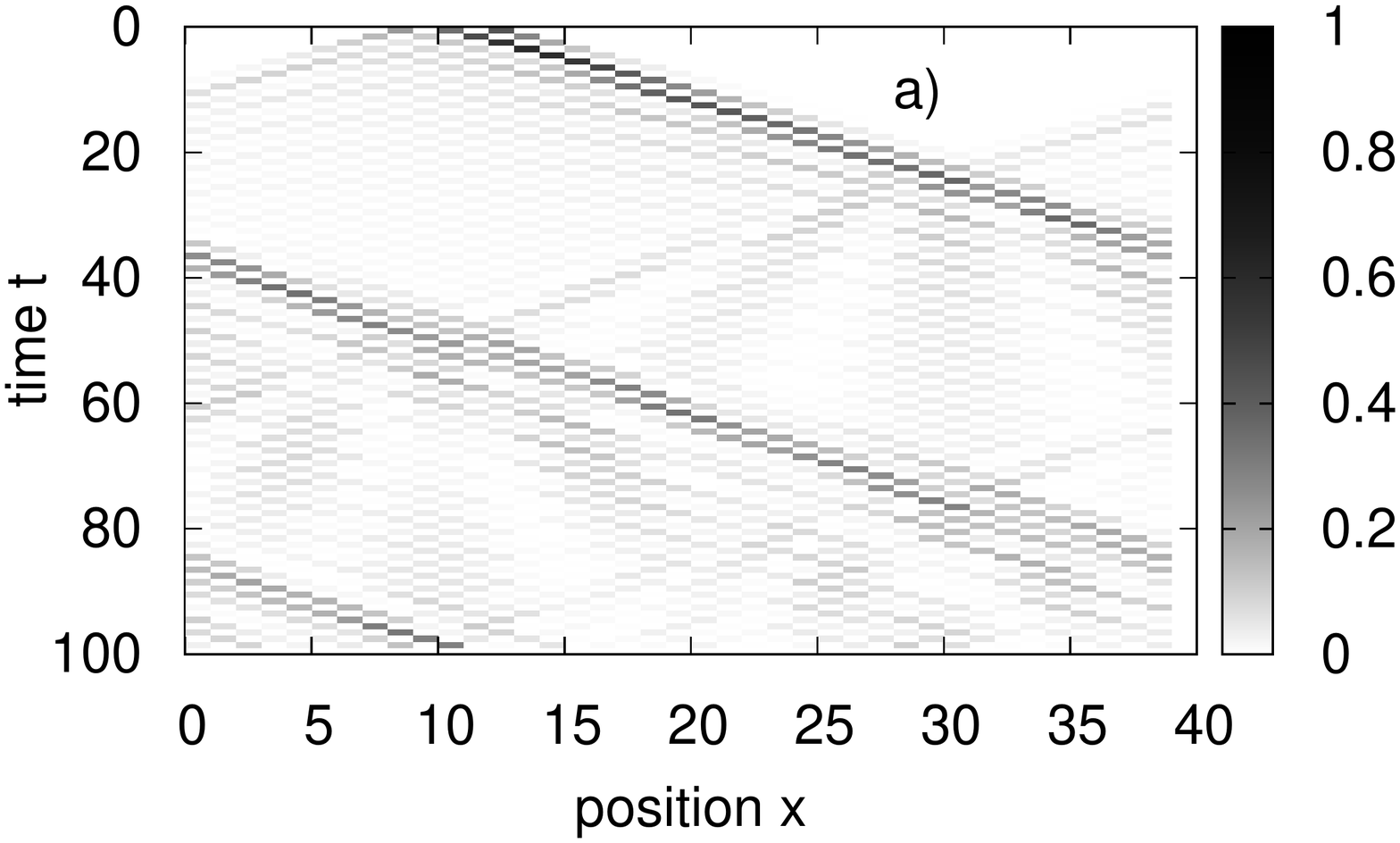}
\includegraphics[width=8cm, trim=3.1cm 4.5cm 1.4cm 4.0cm, clip=true]{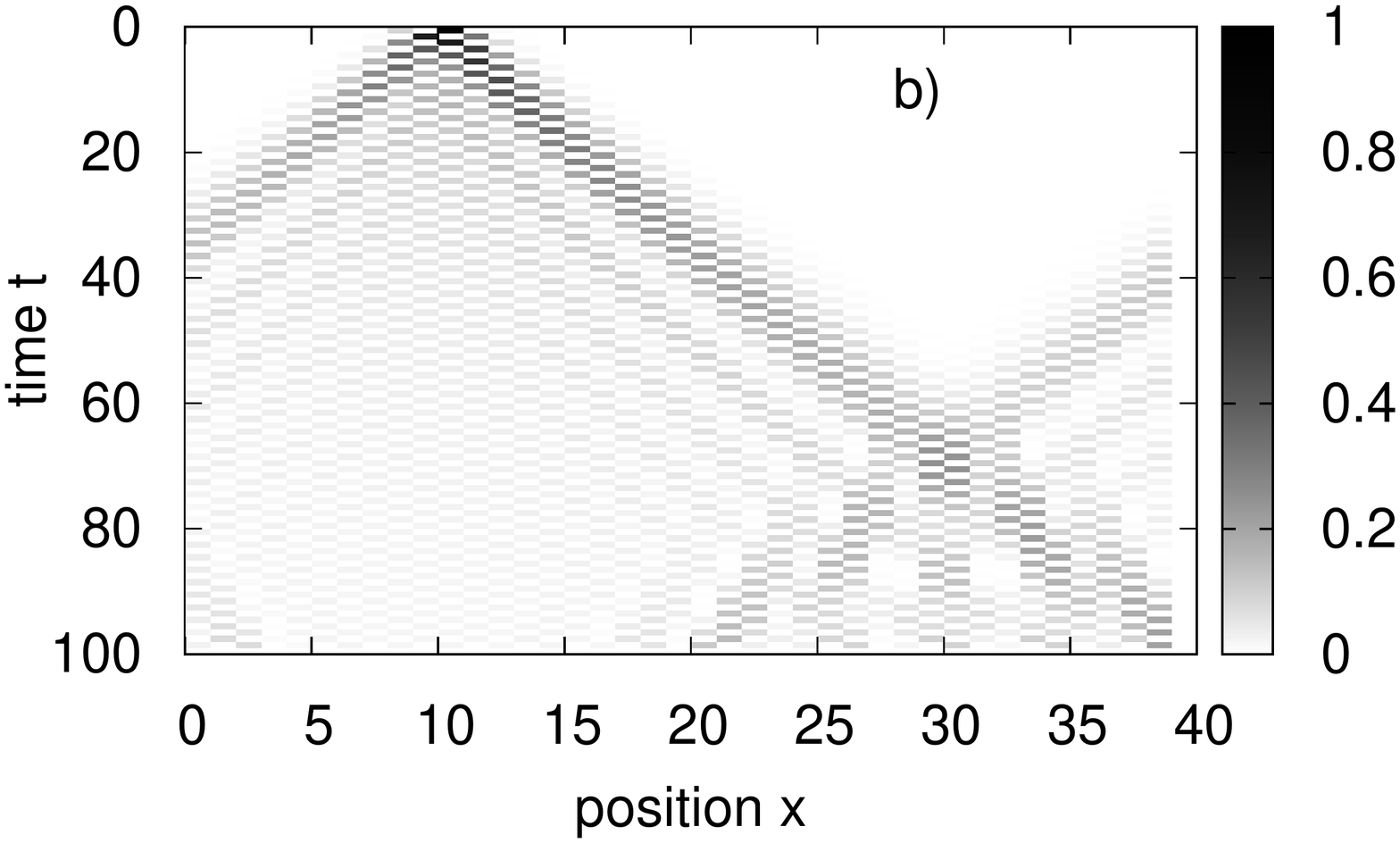}
\includegraphics[width=8cm, trim=3.1cm 4.5cm 1.4cm 4.0cm, clip=true]{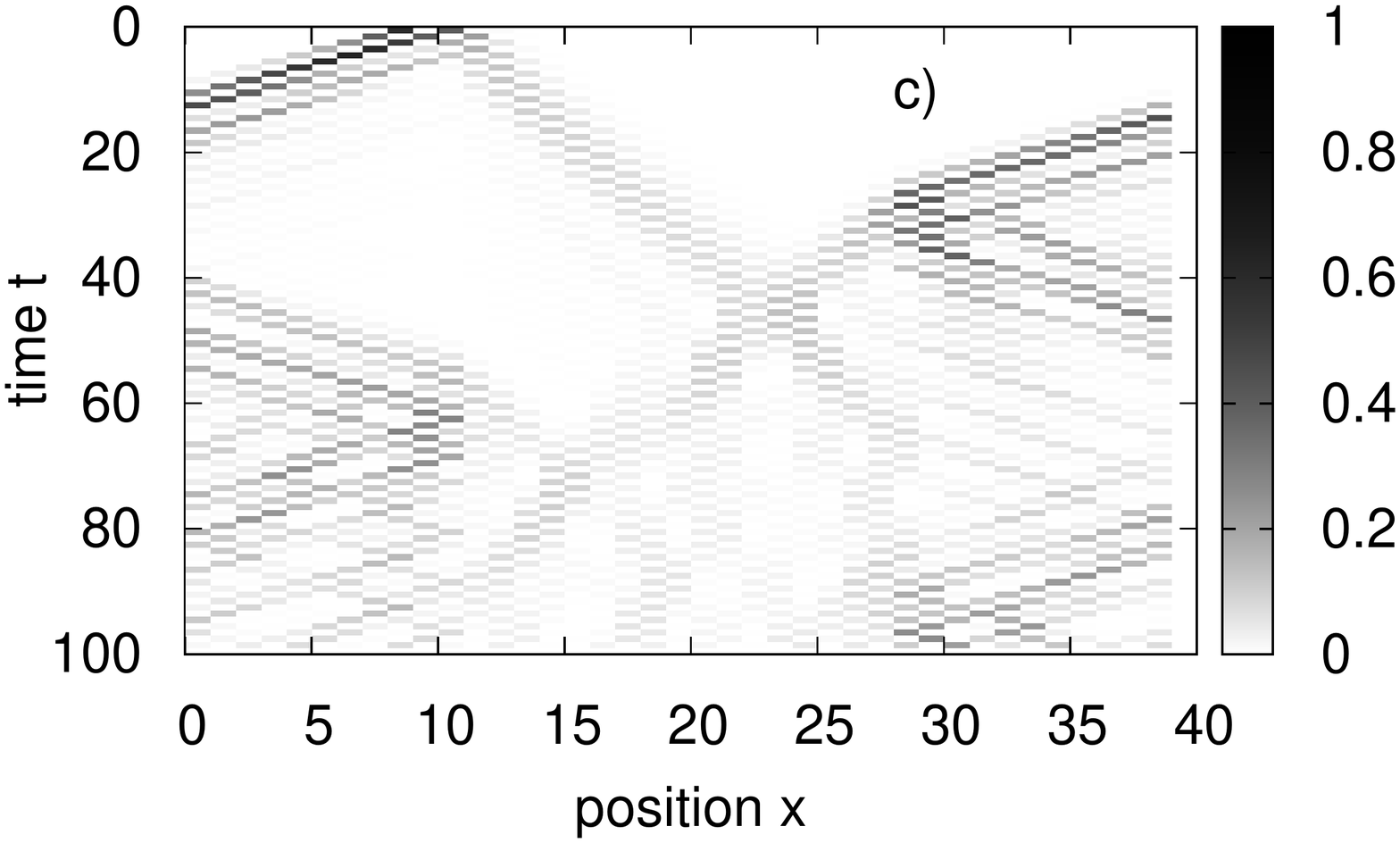}
\includegraphics[width=8cm, trim=3.1cm 4.5cm 1.4cm 4.0cm, clip=true]{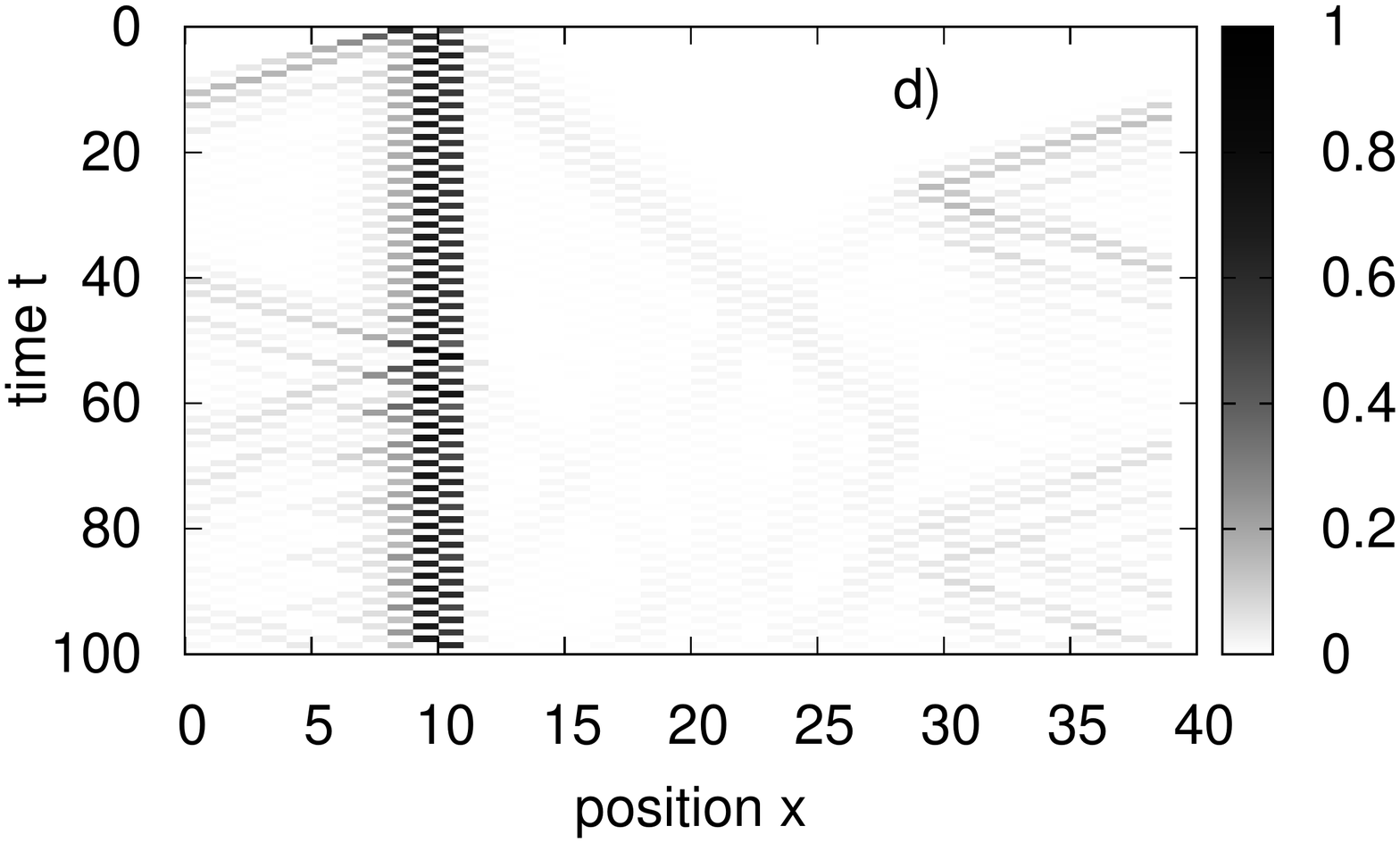}
\caption{Time dependence of the probability
  distribution (color coding corresponding to
  $\abs{\Psi_{x,\uparrow}}^2+\abs{\Psi_{x,\downarrow}}^2$) of a
  walker on a lattice consisting of $N=40$ sites with
  periodic boundary conditions. The coin operator is taken as
  $R(\theta(x))$, with $\theta(x)=\theta_A$ for $10<x<31$ and
  $\theta(x)=\theta_B$ otherwise. In each case we start the walker
  localized on the boundary betwen the domains of $\theta_A$ and
  $\theta_B$, on site $x=10$, with spin up. In cases a) and b), the walk is homogeneous: in
  a) $\theta_A=\theta_B=0.4\pi$, and in b) $\theta_A=\theta_B=0.2
  \pi$.  In c), an inhomogenous system is considered, with a sharp
  boundary between two bulks with the same topology: $\theta_A=0.2
  \pi$ and $\theta_B=0.4 \pi$. In d): $\theta_A=-0.2
  \pi$ and $\theta_B=0.4 \pi$: Here the two domains have different
  topology. Accordingly, a significant part ot the wavefunction of the
  walker gets trapped at the interface. 
} 
\label{fig:map_plots}
\end{figure}

Whether the simple quantum walks with $0<\theta<\pi$ and
$-\pi<\theta<0$ constitute different topological phases can be checked
by considering an inhomogeneous system. As an illustration, we show a
simple choice, a quantum walk on $N=40$ sites, with $U=SR(\theta_A,
\theta_B)$,  where the rotation operator reads 
\begin{align}  
R(\theta_A, \theta_B) &= \sum_{x\in A} \ket{x}\bra{x} \otimes e^{-i\theta_A \sigma_y}
+ \sum_{x\notin A} \ket{x}\bra{x} \otimes e^{-i\theta_B \sigma_y}.
\label{eq:R2_define}
\end{align} 
The domain with rotation angle $\theta_A$ is defined by
\begin{align}
x\in A &\Leftrightarrow 10<x\le 30.
\end{align}
We start the walker localized at $x=10$, with spin
up. As shown in Fig.~\ref{fig:map_plots}, when $\theta_A = \theta_B$,
the walker spreads, with the maximum of the probability spreading with
the maximum of the group velocity. If
$\theta_A$ and $\theta_B$ are different, but in the
same phase, there are diffraction effects at the boundaries $x=10$ and
$x=30$. If
$\theta_B$ and $\theta_A$ have different signs, a part of the walker
is localized at the boundary at $x=10$.

Observed more closely, it is apparent
that the walker trapped at the domain boundary in
Fig.~\ref{fig:map_plots}c) performs a ``zigzag'' motion. 
This is a consequence of the fact that there are not one, but two
localized states at the interface. These are SLS partners of each
other, and thus 1) their energies differ by $E=\pi$ and 2) that their
wavefunctions are related by multiplication by $\tau_z$, the
sublattice symmetry operator defined in Eq.~\eqref{eq:tau_z_def}.
This zigzag motion has already been seen in experiment
\cite{kitagawa_observation}, and its origin in the existence of two
bound states has also been inferred. We now clarify the fact
that it is sublattice symmetry that ensures that bound states always
come in pairs of $E=0$ and $E=\pi$.     

The existence of a pair of topologically protected bound states can be
inferred based on the ``adiabatic continuation'' argument, as, e.g.,
in Kitagawa's pedagogical paper\cite{kitagawa_introduction}. To obtain
a more complete picture, we need to find the topological invariants
associated with the gapped phases.


\subsection{Topological invariants $Q_0$, $Q_\pi$}
Topological invariants for periodically modulated quantum systems have
been suggested by Jiang et al. \cite{akhmerov_majorana}, via an
elaborate construction. For
a periodically driven chain with particle-hole symmetric effective
Hamiltonian, they suggest a $\mathbb{Z}_2\times\mathbb{Z}_2$
topological invariant, $(Q_0,Q_\pi)$. To define the invariant, they
introduce a set of time dependent Hamiltonians
$H_{T} (t)$, with the parameter $0\le T \le 1$ 
specifying the time period. $H_{T} (t)$ should be a
smooth function of $T$, with
\begin{align}
H_{T=0} (t) &= 0; \\
H_{T=1} (t) &= H(t). 
\end{align}
To each $H_{T} (t)$ we can define the corresponding
Floquet operator $U_T = \mathbb{T}e^{-i\int_0^T
  H_T(t) dt}$, and the corresponding effective
Hamiltonian $H_{\mathrm{eff},T}$. In the bulk, $H_{\mathrm{eff},T}$ is translation invariant, and
has a spectrum $E_T^{(n)}(k)$. 

The next step 
to obtain the topological invariant $Q_0$, is to count the parity of the number
of times the gap at $E=0$ closes during the path $T=0\to 1$. Because
of Particle-Hole Symmetry, gaps at $\pm k$ close at the same time. It
is therefore enough to count the number of solutions to
$E_T^{(n)}(k=0)=m \pi$ with arbitrary even integer $m$, add to this
the number of solutions to $E_T^{(n)}(k=\pi)=m \pi$ with arbitrary
even integer $m$, divide the sum by 2 (gaps close from both
directions), and take modulo 2 of the result. Repeating the same
process for arbitrary odd integers $m$ gives us the invariant $Q_\pi$.

For the quantum walk, we can evaluate the topological invariants
$Q_\pi$ and $Q_0$ without following the elaborate construction of
Jiang et al. As shown in Appendix \ref{sec:z2_jiang}, as long as we are only
interested in the \emph{differences} between the topological
invariants of two phases, say $A$ and $B$ -- and this is all that
matters for the physics -- there is a quite straightforward method. 1)
Select any point in parameter space representing a phase $A$ , 2)
connect it via a continuous path in parameter space to a
representative point for phase $B$. 3) Count the parity of the number
of times the gap around $E=0$ closes along this path, to obtain the
invariant $Q_0$, and 4) similarly for the gap around $E=\pm \pi$ to
obtain $Q_\pi$. This construction shows that the gapped phases
$-\pi<\theta<0$ and $0<\theta<\pi$ differ in both invariants $Q_0$ and
$Q_\pi$. This completes the bulk-boundary correspondence picture for
the edge states at the interfaces between these phases.

\section{Quantum walk on a finite line}
\label{sect:finite_line} 

To have a discrete time quantum walk on a finite line, we need to
terminate the 1D lattice. There are
two ways to accomplish this: 1) changing the
coin operators at the boundaries or 2) Cutting the bonds with reflection.

\subsection{Reflective Coin}  
Totally reflective coins have already been considered in the literature. 
Obuse and Kawakami
\cite{obuse_delocalization} mention that $\theta=-\pi/2$
gives a reflective coin with edge states if the bulk has $\theta>0$ and $\theta=+\pi/2$ should be taken
for $\theta<0$. Kitagawa \cite{kitagawa_introduction} explains why
this is so using an adiabatic continuation argument. For the sake of
completeness we briefly summarize a different derivation here. 

The totally reflective unitary coin operator reads 
\begin{align}
R_0 = \begin{pmatrix}
0 & e^{i\phi}\\
e^{i\xi} & 0
\end{pmatrix}.
\end{align}
To keep Particle-Hole Symmetry represented by complex conjugation, we
would like to choose $R_0$ to have only real elements. That leaves us
4 choices for $R_0$: $\pm \sigma_x$ and $\pm i \sigma_y$. A walker
only sees the totally reflective coin from one side, and thus we can
take $R=\pm i\sigma_y$ without loss of generality. This corresponds to
choosing the reflective coin to have a rotation angle $\theta$ which
is in the middle of one of the gapped phases. If this is the same
gapped phase as that of the bulk, there are no bound states at the
boundary. If it is not the same as that of the bulk, there are two
bound states with energies 0 and $\pi$, that are sublattice partners
of each other.

\subsection{Cutting a link} 

Unitarity of the quantum walk is a strong constraint on how we
can cut a link. When the walker attempts to jump over a link that is
cut, it has to end up in a state which is unaccessible to it from any
other state. The only states that are ``not taken'' are those to
either sides of a cut link. Therefore, the only option to implement a
totally cut link, is to introduce a spin flip instead of a jump. It is
still possible to include a phase shift along with the spin
flip.
To retain PHS, this phase shift can only be chosen to be $\pm 1$. 
In much the same way as with the reflective coin above, without loss
of generality, we can
fix a phase of $-1$ upon reflection from one of the sides. Cutting the
link between sites $y$ and $y+1$ is implemented by altering the 
shift operator $S$:
\begin{align} 
\label{eq:SC_def}
 S_{(y)}&= \sum_{x\neq y} S_{x,x+1} \pm C_{y,y+1}.
\end{align}
Here, the shift operators for the ``link'' and ``cut link'' between
sites $x$ and $x+1$ are defined as 
\begin{align}
  S_{x,x+1} &= \ket{x,\downarrow} \bra{x+1,\downarrow} + \ket{x+1,\uparrow} \bra{x,\uparrow};\\   
  C_{x,x+1} &= \ket{x+1,\uparrow}\bra{x+1,\downarrow} - \ket{x,\downarrow} \bra{x,\uparrow}.
\end{align} 
The $\pm$ in Eq.~\eqref{eq:SC_def} represents the choice of the
reflection phases allowed by Particle-Hole symmetry.

\subsection{Partially cut links in the bulk} 
In order to use bulk-boundary correspondence, we need to connect the ``cut link'' to
the ``uncut link'' by way of a continuous parameter in the Floquet
operator. The first idea here, the introduction of an additional
``link rotation angle'' $\phi$, works:
\begin{align}
  S_{x,x+1}(\phi) &= \cos (\phi) \, S_{x,x+1} + \sin (\phi)\, C_{x,x+1}.
\end{align}
In the bulk, this is equivalent to the ``split-step'' walk of Kitagawa et al
\cite{kitagawa_exploring},  where the spin-$z$ dependent displacement 
is broken down to two successive steps: 
\begin{align} 
S_\downarrow &= \sum_{x=1}^{N}\big( \ket{x-1}\bra{x} \otimes \ket\dd\bra\dd + 
\ket{x}\bra{x} \otimes \ket\uu\bra\uu \big); \\ 
 S_\uparrow &= \sum_{x=1}^{N}\big( \ket{x}\bra{x} \otimes \ket\dd\bra\dd + 
\ket{x+1}\bra{x} \otimes \ket\uu\bra\uu \big); 
\end{align}
\begin{align}
  U_2(\theta, \phi) &= S(\phi) R(\theta) = S_\uparrow e^{-i\phi
    \sigma_y} S_\downarrow e^{-i\theta \sigma_y}.
\label{eq:splitstep_u}
\end{align} 
As shown in \cite{kitagawa_exploring}, $S(\phi=0)=S_\uparrow
S_\downarrow = S$. 

With a partially cut link, the sublattice symmetry of the Floquet
operator $U$ is broken. A walker that is reflected off an edge has the
same $x$ at the end of the timestep as at the beginning (and possibly
even the same spin), therefore the graph of $U$ cannot be
bipartite. Therefore, the gaps at $E=0$ and $E=\pm \pi$ now can open
and close independently (and the energy $E=\pm \pi/2$ is no longer
protected by symmetries).

\subsection{Topological phases of the split-step walk}  


\begin{figure}
\includegraphics[width=8cm]{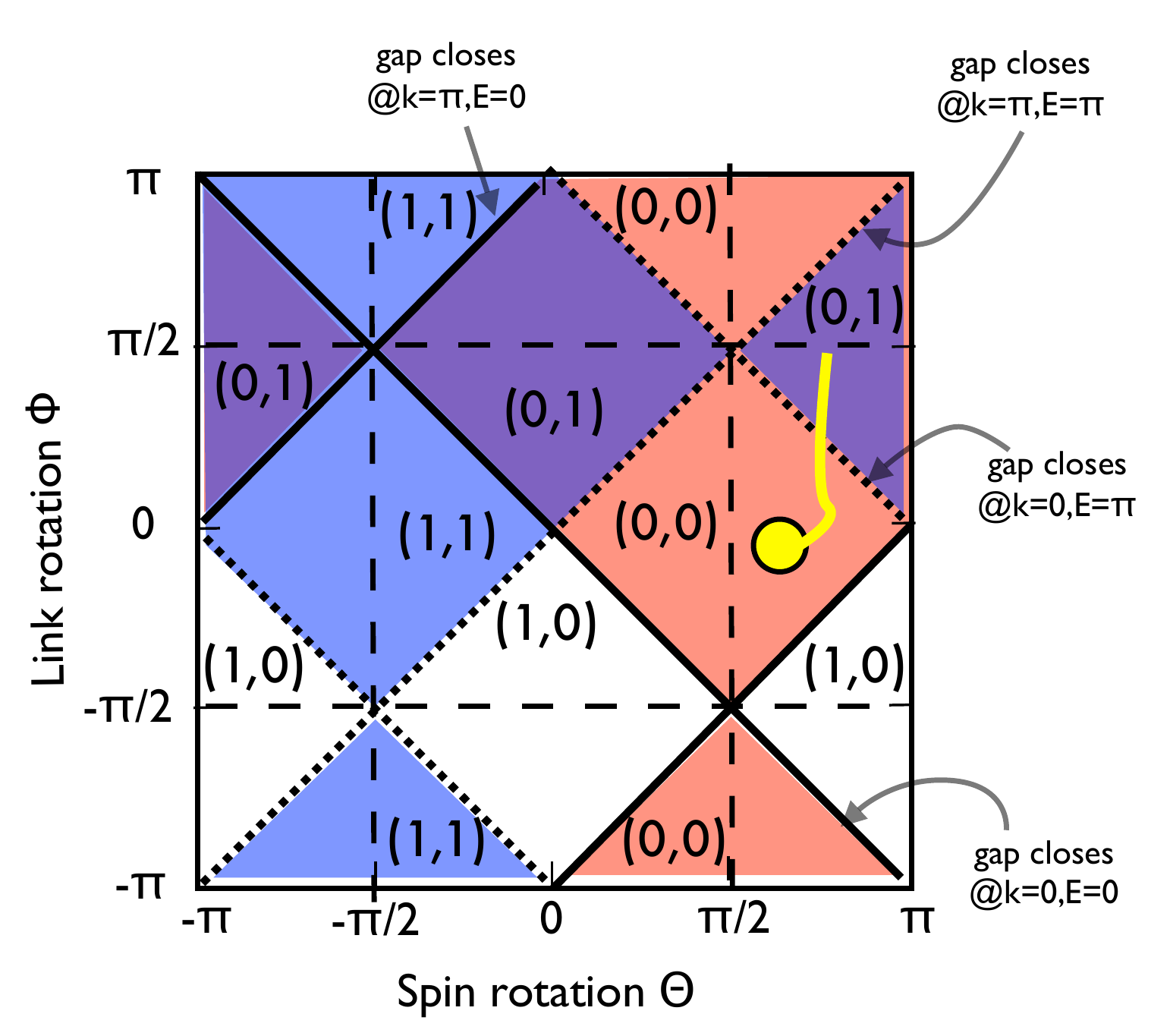}
\caption{Phase map of a 1D quantum walk with partially cut links. The
  various gapped domains (different shadings) have a
  $\mathbb{Z}_2\times \mathbb{Z}_2$ invariant associated with them,
  which is indicated as a pair of numbers $(Q_0, Q_\pi)$. Separating
  these domains are the lines where the gap at $E=0$
 closes (continuous line), or the 
gap at $E=\pi$ closes (dotted line).   closes. The vertical (horizontal) slashed lines denote the parameters
  corresponding to reflective coins (cut links). To find the number of
  edge states at an edge with cut links, a the point corresponding to
  the quantum walk (circle) is connected to the horizontal slashed
  line representing the boundary conditions. In the example shown,
  there is a single bound state with $E=\pi$.  
} 
\label{fig:map_cut_links}
\end{figure}

 The split-step quantum walk, Eq.~\eqref{eq:splitstep_u}, has two
parameters, the ``coin angle'' $\theta$ and the ``bond angle''
$\phi$. The parameter space is therefore now a torus. The Floquet
timestep operator $U_2$ reads  
\begin{align} 
U_2(k) &= e^{-i\sigma_z k/2} e^{-i\sigma_y \phi}
e^{-i\sigma_z k/2} e^{-i\sigma_y \theta}.
\label{eq:U2_def}
 \end{align}  
The dispersion relation is straightforwardly derived\cite{kitagawa_exploring},
\begin{align}  
\cos E(k) &= \cos (k) \cos (\theta) \cos (\phi) - \sin (\theta) \sin (\phi)
\end{align}
As can be seen from this dispersion relation,
the split-step quantum walk for generic $\phi$ and $\theta$ has gaps
around $E=0$ and $E=\pm \pi$. The Floquet operator takes on a very
simple form at the time-reversal invariant momenta $k=0,\pi$:
\begin{align}
U_2(k=0) &= e^{-i H(k=0)} = e^{-i\sigma_y (\theta+\phi)};\\
U_2(k=\pi) &= e^{-i H(k=\pi)} = -e^{-i\sigma_y (\theta-\phi)};
\end{align}
Therefore the gap around $E=0$ closes at $k=0$, $\phi=-\theta$, and
$k=\pi$, $\phi=\pm\pi+\theta$, and the gap around $E=\pm\pi$ closes at
$k=\pi$, $\phi=\theta$, and $k=0$, $\phi=\pm\pi-\theta$. 

The parameter space $(\theta,\phi)$ is divided into
4 different gapped topological phases, with topological invariants $Q_0$ and
$Q_\pi$, as shown in Fig.~\ref{fig:map_cut_links}. Selecting
$\theta=\pi/2, \phi=0$ as a reference point, we define the values
of the invariants for the domain around this point as 
$(0,0)$. For any point in parameter space, we 1) pick a continuous
path in parameter space connecting it with the reference point, 2)
count the parity of the number of times gap around $E=0$ ($E=\pi$)
closes along the path. The parities give the values of the invariant
$Q_0$ ($Q_\pi$). Because of Particle-Hole Symmetry, it is enough to
count the gap closings at the time-reversal invariant momenta $k=0$ and $k=\pm\pi$.

\begin{figure}
\includegraphics[width=8cm]{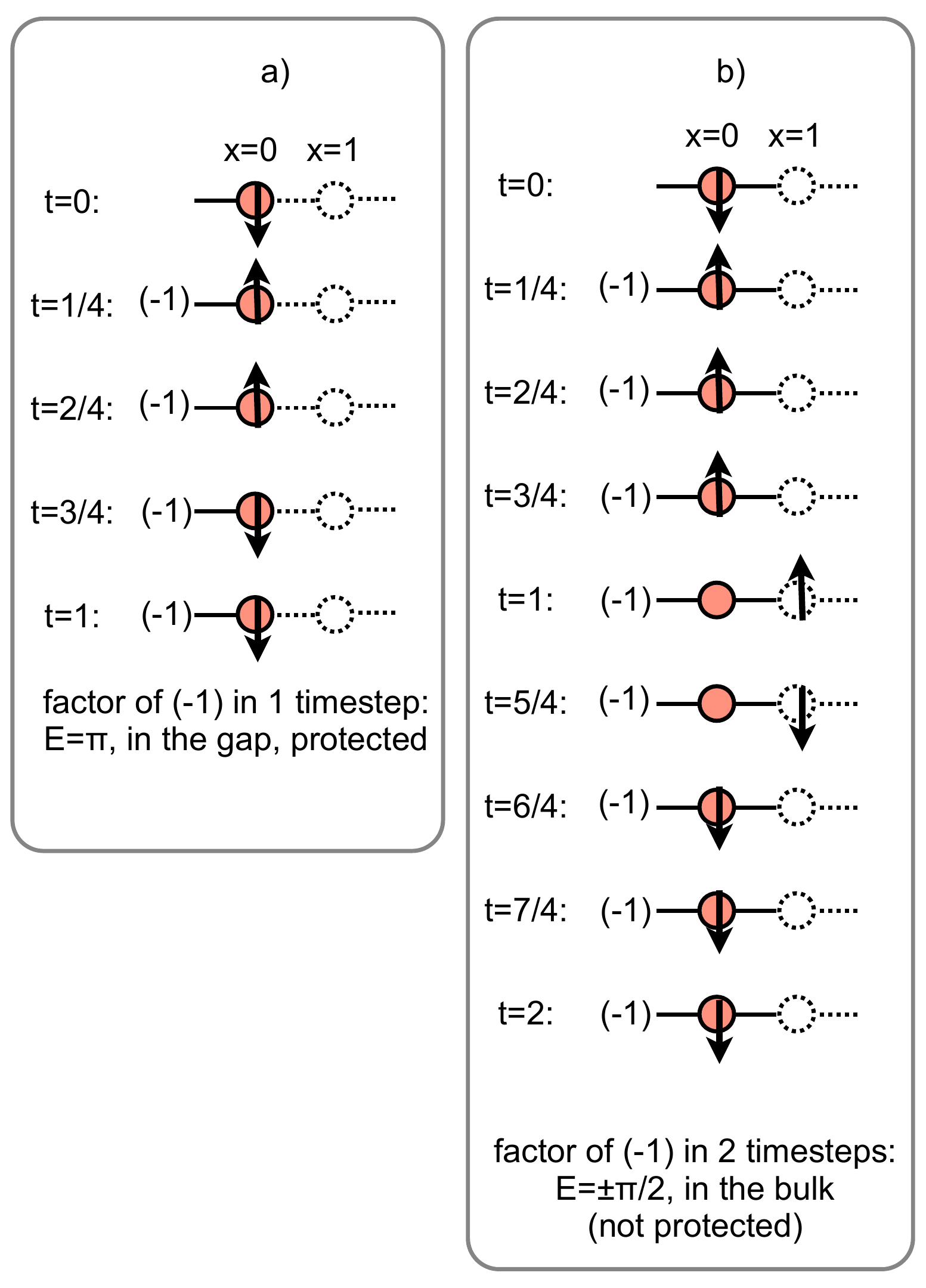}
\caption{Timesteps for a walker started from an edge. Contionuous (dotted) circles and
  lines correspond to the sites and links of the bulk (boundary). The
  timesteps are broken down to
4 successive operations, as in Eq.\eqref{eq:U2_def}, each occurring in $1/4$ time. Continuous (dotted)
  circles and lines correspond to the sites and links of the bulk
  (boundary). For simplicity, the bulk is taken with $\theta=\pi/2$,
  and $\phi=0$: a simple quantum walk. The boundary has
  $\theta=\pi/2$, and cut links: $\phi=\pi/2$. If the reflection is
  on a cut link (a), there is a protected midgap edge state with
  energy $\pi$. If the reflection happens on a reflective coin (b),
  during two timesteps, the walker acquires a phase of
  (-1). Superpositions of the states at $t=0$ and $t=1$ with a
  relative phase of $i$ ($-i$) are therefore stationary states with
  energy $-\pi/2$ ($\pi/2$), not protected by particle-hole symmetry.
}
\label{fig:cut_links_ambiguous}
\end{figure}

Setting $\phi=0$ corresponds to the original ``simple'' quantum
walk. Setting $\phi=\pm \pi/2$ corresponds to two different ways in
which the bonds can be cut in a unitary and particle--hole-symmetric
way. As illustrated in  Fig.\ref{fig:map_cut_links}, using the bulk-boundary correspondence, we find that
for a generic quantum walk with $\phi=0$, each edge defined by cutting
a link in a particle-hole-symmetric way hosts a single topologically
protected edge state.
Whether the energy of that state is $E=0$ or
$E=\pi$ depends on the bulk quantum walk and on the way in which the
link is cut (on the reflection phase). 
In the example of \ref{fig:map_cut_links}, we find that the energy of
the bound state is $E=\pi$. Note that this is independent of
the path itself, and also of its endpoint on the line representing the
cut links - except if this endpoint is at $\theta=\pm \pi/2$, in which
case the details of the edge need to be specified to show whether the
reflection is off of the reflective coin or from the cut link. This is
illustrated in Fig.~\ref{fig:cut_links_ambiguous}.    

\subsection{Edge states between two bulks with the same Floquet operator}

The topological invariant for a quantum walk cannot be inferred from
its effective Hamiltonian alone. Evidence for this has already been noted by
Kitagawa\cite{kitagawa_introduction}, who describes pairs of bound states
between topological phases with the same ``winding number''. 
The most striking illustration of
this statement, however, is a pair of edge states between two bulks with the
same Floquet operator.  

Consider an inhomogeneous quantum walk with periodic boundary
conditions, consisting of two bulks, separated by a sharp
boundary. The dynamics is given by the split-step protocol, and the
bulks differ in both parameters $\theta$ and $\phi$:
\begin{align} 
 U_2 = S_\uparrow R(\phi_A,\phi_B) S_\downarrow R(\theta_A,\theta_B),
\label{eq:U2_inhom}
 \end{align}
with the inhomogeneous rotation operator $R(\theta_A, \theta_B)$
defined as in Eq.~\eqref{eq:R2_define}. Taking $\phi_B = \phi_A+\pi$
and $\theta_B=\theta_A+\pi$, the translationally
invariant bulk time evolution operators of the two domains read 
\begin{align} 
 U_A &= S_\uparrow e^{-i\phi_A \sigma_y} S_\downarrow e^{-i\theta_A \sigma_y} \\
 U_B &= S_\uparrow e^{-i(\phi_A+\pi) \sigma_y} S_\downarrow 
 e^{-i(\theta_A+\pi) \sigma_y}. 
\end{align} 
Note that since $e^{-i\pi \sigma_y} = -1$, we have 
\begin{align}
  U_A &= U_B.  
\end{align}
As can be seen from the phase map,  Fig.\ref{fig:map_cut_links}, the
simplest path in the parameter space connecting two such points
intersects gap closings at $E=0$ and at $E=\pi$ once. Thus, there  are 2 edge states between
these two bulks, with energies $0$ and $\pi$. 

\begin{figure}
\includegraphics[width=8cm]{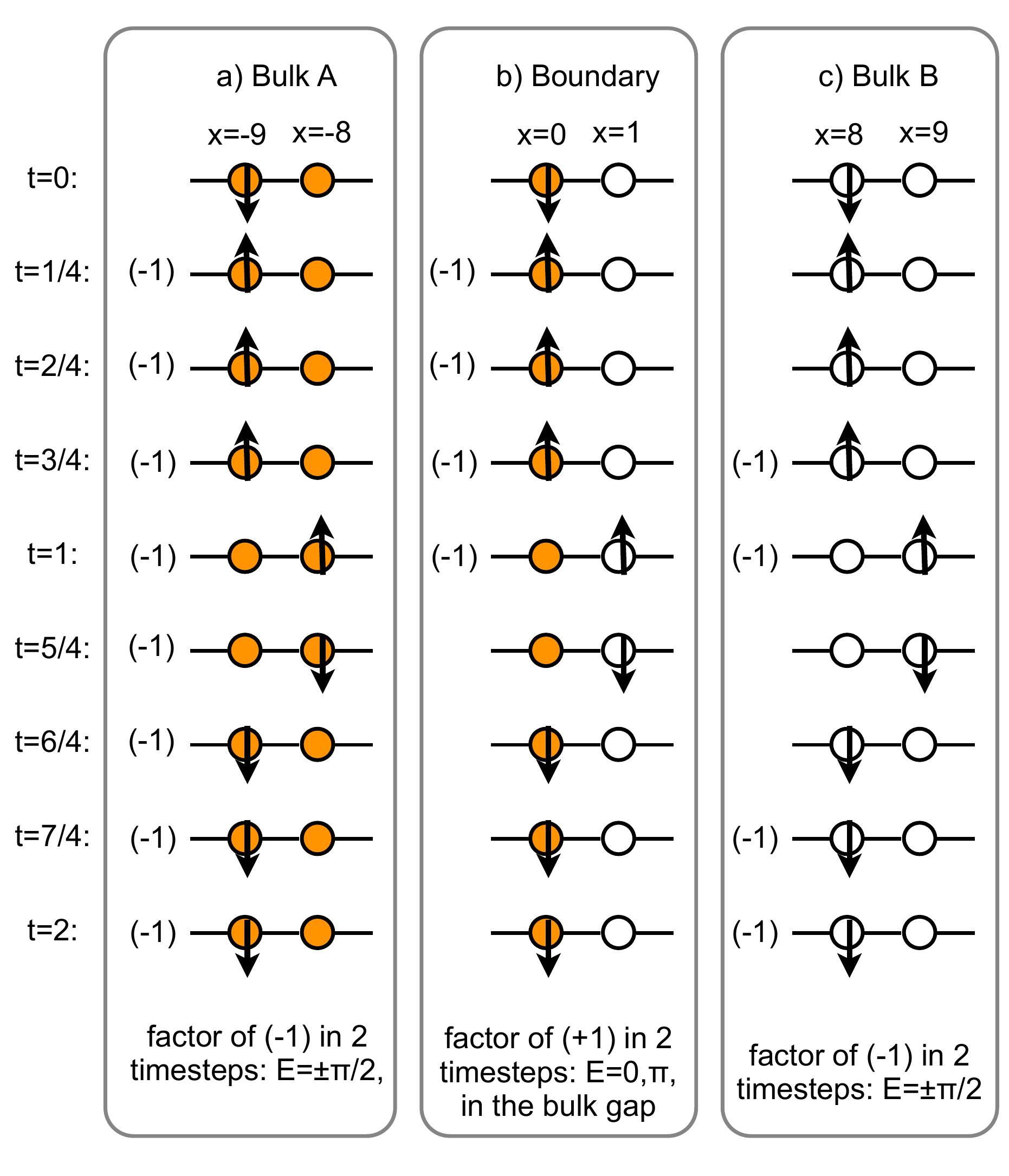}
\caption{Two successive timesteps of a quantum walk, with a walker
  started in bulk A (a), at a sharp boundary (b), or in bulk B (c). 
Each timestep is broken down to its 4 stages,
given by the 4 factors in
 $U_2(k) = S_\uparrow R(0,\pi) S_\downarrow R(\pi/2,-\pi/2)$,
with $R(\theta_A,\theta_B)$ as defined in Eq.\eqref{eq:R2_define},
with $x\in A \leftrightarrow x<1$. 
In each case, the walker
returns to its initial site after 2 timesteps. In the bulk,
during the 2 timesteps a phase factor of (-1) is acquired by the
walker, showing that stationary states (superpositions of the states
at $t=0$ and $t=1$ with relative phase $\pm i$) have quasienergy $\mp \pi$. 
At the boundary, this factor is (+1), therefore even and odd
superpositions of the states at $t=0$ and $t=1$ are stationary states
with energy $0,\pi$. These are at the
topologically protected midgap states. 
} 
\label{fig:boundary_floquet}
\end{figure}

Perhaps the simplest concrete example is a boundary between $\phi=0,
\theta=\pi/2$, and $\phi=\pi, \theta = -\pi/2$. 
We illustrate this in Fig.~\ref{fig:boundary_floquet}. 

\section{Conclusions}

In this paper we revisited the topological phases of the 1-dimensional
quantum walk. 
To begin, we identified the symmetries of
the corresponding effective Hamiltonian. In contrast with the
literature, we find that the Hamiltonian belongs to class D, i.e.,
it has a Particle-Hole Symmetry that squares to 1, and no other
symmetries. We argue that the property of the homogeneous quantum walk identified as
Chiral Symmetry should not rightfully be regarded as a symmetry, since
the operator representing it varies from phase to phase.    
We also find that there is an additional symmetry of the timestep
operator, which could be called ``sublattice symmetry'', however,
it does not have any special effect on the topological properties of
the system. 

To identify the topological phases of 1-dimensional discrete-time
quantum walks, however, the bulk effective Hamiltonian (or indeed, the bulk
Floquet operator) is not enough. We have found that a more complete
specification of the experimental realization is needed, e.g.,
the sequences of rotation-translation. We have evaluated
the topological invariant of Jiang et al.\cite{akhmerov_majorana} for
such specifications corresponding to the simple
discrete-time quantum walk, and for the ``split-step'' walk introduced
by Kitagawa et al.\cite{kitagawa_exploring}. For the simple walk we
find two different phases, whose boundary hosts a pair of topologically
protected edge states.  For the split-step walk, we find all 4
different topological phases corresponding to the
$\mathbb{Z}_2\times\mathbb{Z}_2$. We provide a blatant proof of the fact that the
bulk Floquet operator does not contain all the information about the
topological phase: A pair of topologically protected edge states
between two bulks which differ in their experimental description, but
have the same Floquet operator.

The use of periodically modulated external fields to alter the
topological properties of Hamiltonians has been considered by several
authors \cite{floquet_topological,dora_optical}. In all cases, however, these works employ the
same topological invariants as for time-independent systems. It would
be interesting to explore what the complete topological invariant in
these cases is, and under what conditions does it give rise to edge
states that are unique to periodically driven systems. Kitagawa et
al.\cite{kitagawa_periodic} have already shown that for a periodically
modulated hexagonal lattice, edge states can arise between bulk phases
with the same Chern number. However, even for this specific system,
the bulk topological invariant has not yet been defined.

This work was supported by the Hungarian Academy of Sciences (Lend\"ulet Program, LP2011-016). 
We
acknowledge useful discussions with Anton Akhmerov, Tam\'as Kiss, and
Zolt\'an Kurucz. 
\appendix

\section{Sublattice symmetry and the doubling of states}

Any stationary state $\ket{\Psi}$ of a quantum walk with sublattice
symmetry must have support on both sublattices $A$ and $B$. Using the obvious
notation for the projection of a state on a sublattice, 
$\ket{\Psi_{A}} \equiv \underset{j\in A}{\sum} \ket{j}\bra{j}
\ket{\Psi}$, and similarly, 
$\ket{\Psi_{B}} \equiv \underset{j\in B}{\sum} \ket{j}\bra{j} \ket{\Psi}$,
 we have
\begin{align}
  \ket{\Psi} &= \ket{\Psi_A}+\ket{\Psi_B}.
\end{align}
For stationary states, $U \ket\Psi = e^{-iE} \ket\Psi$, which gives us $U
\ket{\Psi_{A,B}} = e^{-iE} \ket{\Psi_{B,A}}$. 
Therefore, both $\ket{\Psi_A}$ and $\ket{\Psi_B}$ are eigenstates of
the step-doubled walk,
\begin{align}
  U^2 \ket{\Psi_{A,B}} &= e^{-2iE} \ket{\Psi_{A,B}} . 
\end{align} 
Doubling the timestep gives a walk on only one sublattice, since
$U^2\tau_z = \tau_z \tau_z U \tau_z \tau_z U \tau_z = \tau_z U^2$, and
projection to sublattice $A,B$ is given by $1/2 (1\pm \tau_z)$.
Therefore, we can double the timestep and restrict to sublattice
$A$. For any eigenstate of $U^2$ with energy $E_2$, we have :
\begin{align*}
  U\ket{\Psi_{A}} &= \ket{\Psi_B};\\
  U\ket{\Psi_{B}} &= e^{-iE_2} \ket{\Psi_A};
\end{align*}
Introducing $E = E_2/2$, we can reconstruct the two 
eigenstates of $U$ with energies $E_2$ and $E_2+\pi$, linked by
sublattice symmetry:   
\begin{align*}
  U (\ket{\Psi_A} \pm e^{-iE} \ket{\Psi_B}) 
  &= \pm e^{-iE} (\ket\Psi_A \pm e^{-iE} \ket{\Psi_B}).
\end{align*}
Therefore, any energy eigenstate of $U^2$, projected onto one of the
sublattices, gives us two energy eigenstates of $U$, related to each
other by the sublattice symmetry.  This means that we can double the
timestep without losing any energy eigenstates.

\section{The $\mathbb{Z}_2 \times \mathbb{Z}_2$ invariant in parameter
space}
\label{sec:z2_jiang} 
To infer the number of topologically protected edge states at an edge
between two bulks $A$ and $B$, i.e., to apply the bulk-boundary
correspondence, we do not need to know the values of the topological
invariants $(Q_0,Q_\pi)$ in these bulks. It is enough to know the
amounts by which the values of these invariants change between the two
bulks. Therefore it is not necessary to find the complete set
$H_{T} (t)$, corresponding to a continuous path in
parameter space to ``doing nothing''. 

We assume two things. First, that a set of experimental Hamiltonians exists
for bulk $A$ that connects it to ``doing nothing'': $H^{A}_{T} (t)$, with
$H^{A}_{T=0} = 1$ and $H^{A}_{T=1} (t) =
H^{A}_{\mathrm{exp}}$. Second, that for the continuous path in the space of
parameters of the quantum walk, $\theta(x)$, with
$\theta(x=0)=\theta_A$ and $\theta(x=1)=\theta_B$, the experimental
Hamiltonians $H^\theta(x)_\mathrm{exp}$ along the path are also
continuous functions of $x$.   

We construct the path $H^B_{\mathrm{exp},T}$ in the following way:
\begin{align}
H^B_{\mathrm{exp},T}(t) &= H^{A}_{\mathrm{exp},2 T}(2t) \quad \mathrm{if}\quad T<1/2;\\
H^A_{\mathrm{exp},T}(t) &= H^{\theta(2T-1)}_{\mathrm{exp}}(t/T) \quad
\mathrm{if}\quad
T\ge 1/2.
\end{align} 
For this construction, the difference in the invariant $Q_0$ between the bulks
$B$ and $A$ can be obtained by just counting
the number of times the gap around $E=0$ closes along the path
$\theta(x)$. The analogous recipe holds for the invariant $Q_\pi$,
with the gap around $E=\pm \pi$.





\bibliography{walkbib}{}

\end{document}